\documentclass[aip,pop,amsmath,amssymb,reprint]{revtex4-2}

\usepackage{graphicx}            
\usepackage{dcolumn}             
\usepackage{bm}                  

\usepackage[version=4]{mhchem}
\usepackage{booktabs}
\usepackage{multirow}
\usepackage{fix-cm}
\usepackage{microtype}
\usepackage[colorlinks=true, linkcolor=blue, citecolor=blue, urlcolor=blue, breaklinks]{hyperref}

\begin{document}

\preprint{AIP/123-QED}

\title[Influence of gas flow rate on modes of reactive oxygen and nitrogen species in a grid-type surface DBD]{Influence of gas flow rate on modes of reactive oxygen and nitrogen species in a grid-type surface dielectric barrier discharge}

\author{\vspace{0.25em}A. N. Torres Segura}
\email{torressegura@aept.rub.de}
\affiliation{ 
    Chair of Applied Electrodynamics and Plasma Technology, Faculty of Electrical
    Engineering and Information Technology, Ruhr University Bochum, Germany
}

\author{K. Ikuse}
\affiliation{ 
    Hamaguchi Laboratory, Division of Materials and Manufacturing Science, Graduate School of Engineering, Osaka University, Japan
}

\author{S. Hamaguchi}
\affiliation{ 
    Hamaguchi Laboratory, Division of Materials and Manufacturing Science, Graduate School of Engineering, Osaka University, Japan
}

\author{A. R. Gibson}
\affiliation{  
    Research Group for Biomedical Plasma Technology, Faculty of Electrical Engineering and Information Technology, Ruhr University Bochum, Germany
}
\affiliation{  
    York Plasma Institute, School of Physics, Engineering and Technology, University of York, United Kingdom
}

\author{L. Schücke}
\email{schuecke@aept.rub.de}
\affiliation{ 
    Chair of Applied Electrodynamics and Plasma Technology, Faculty of Electrical
    Engineering and Information Technology, Ruhr University Bochum, Germany
}
\affiliation{  
    Hamaguchi Laboratory, Division of Materials and Manufacturing Science, Graduate School of Engineering, Osaka University, Japan
}
\affiliation{  
    Research Group for Biomedical Plasma Technology, Faculty of Electrical Engineering and Information Technology, Ruhr University Bochum, Germany
}

\date{\today}

\begin{abstract}
The presented work investigates a surface dielectric barrier discharge (SDBD) operated dry synthetic air as the working gas using a combination of experimental measurements and simulations. The primary objective is to characterize the production and consumption dynamics of reactive oxygen and nitrogen species to enhance the understanding of their formation and facilitate control of the discharge for applications. Densities of O\textsubscript{3}, NO\textsubscript{2}, and N\textsubscript{2}O\textsubscript{5} are measured under varying gas flow rates, utilizing optical absorption spectroscopy as the diagnostic method. A semi-empirical chemical kinetics model is developed based on a compilation of reactions from previous studies on similar types of discharges. The results reveal two previously known and distinct operating modes, with a mode transition occurring between the modes as the flow rate is varied. The results indicate the dependency of the mode transition on the density of sufficiently vibrationally excited nitrogen molecules, which is represented in the model by an increased vibrational temperature at lower gas flow rates. Furthermore, key reactions responsible for the production and consumption of ozone and nitrogen oxides are identified, providing insight into the importance of macroscopic parameters, such as gas temperatures and different time constants, that influence the nonlinear balance of these reactions.
\end{abstract}

\keywords{dielectric barrier discharge, reactive oxygen and nitrogen species, plasma chemistry, optical absorption spectroscopy, chemical kinetics simulation}

\maketitle

\section{Introduction}
\label{chap:introduction}

Atmospheric pressure plasma sources are widely used and researched, for a range of applications such as biomedicine \cite{woedtke2022}, sustainable production of basic chemicals \cite{sun2024_review}, and surface processing \cite{kim2003}. For these applications a range of different types of plasma sources are available, such as the dielectric barrier discharge (DBD) \cite{subedi2017}, the plasma jet \cite{schutze1998}, or the plasma needle \cite{kieft2004}. In this work the focus will be on DBDs, where the plasma discharge occurs between two electrodes, with at least one of them covered by a dielectric material. This dielectric material is crucial to prevent the formation of a thermal arc, as it prevents the continuous flow of current, thus keeping the plasma in a controlled non-thermal state. \\

There are two main classifications of DBDs, the volume dielectric barrier discharge (VDBD) and the surface dielectric barrier discharge (SDBD). In the VDBD configuration one or both electrodes are covered by a dielectric material, or the dielectric layer may be located within the electrode gap containing the discharge gas. A key feature of the VDBD configuration is that the volume of gas between the electrodes is occupied by the plasma, which allows adjusting and controlling the volume of the discharge by varying the size of the dielectric and electrode areas. On the other hand, in the case of the SDBD configuration, the space between the electrodes is typically completely occupied by the dielectric material, which forces the plasma discharge to occur along the surface of the dielectric material. Some applications typical for this configuration are the treatment of volatile organic compounds VOCs~\cite{boeddecker2024, schuecke2020, schmidt2015}, plasma actuators in aerodynamic systems~\cite{yang2015, zhu2013}, or generation of ozone and other reactive species~\cite{simek2012}. The use of SDBDs for the controlled generation of ozone, as well as nitrogen oxides from air as a process gas, enables a range of potential applications. For example, ozone O\textsubscript{3} is found to be very useful for wastewater treatment \cite{rice1996} and sterilization, while nitrogen oxides NO\textsubscript{x} have application for plant growth stimulation \cite{sirova2011} and wound healing~\cite{anatoly2005}. \\

An important feature of reactive species production in SDBDs is that they exhibit mode transitions, where the densities of reactive species can vary over several orders of magnitude. The major modes for DBDs operated in air are the O\textsubscript{3} dominated mode and the NO\textsubscript{x} dominated mode, as observed and studied in \cite{shimizu2012, pavlovich2014, park2018, liu2022, xi2022, zhu2023, huh2024, sun2024}. In many SDBD systems the transition from the O\textsubscript{3} dominated mode to the NO\textsubscript{x} mode occurs after the discharge has been active for a certain period of time, typically in the range of tens of seconds. Shimizu~\textit{et al.} \cite{shimizu2012} and Park~\textit{et al.}~\cite{park2018} have developed semi-empirical models to explain the mode transition on the basis of the temporal variation in the vibrational excitation of N\textsubscript{2} molecules. In these models, the vibrational excitation of N\textsubscript{2} increases exponentially with a certain time constant, leading to NO formation via the reaction below, which generally requires vibrational states above a certain level to occur at a high rate:
\begin{equation}
    \mathrm{O} + \mathrm{N_2(\textit{v})} \rightarrow \mathrm{NO} + \mathrm{N}
\end{equation}

The produced NO subsequently reacts to consume O\textsubscript{3}, and drives the formation of a range of NO\textsubscript{x} species. As such, the mode of operation of an SDBD is of profound importance when it comes to application outcomes, as these are driven by the specific reactive species produced in the system. While the models developed by Shimizu~\textit{et al.} and Park~\textit{et al.} are not self-consistent, in that they require a number of assumptions about the densities of reactive species and vibrational states as input, variation of these free parameters in the model can be used to fit modeled density profiles to those measured experimentally, yielding insights into some of the underlying chemical kinetics. Further, the mode of operation of SDBDs has been demonstrated to depend on the operating conditions, such as the characteristics of the voltage waveform and the power deposited in the discharge \cite{shimizu2012, pavlovich2014, liu2022, zhu2023, sun2024}, and the composition of the feed gas \cite{xi2022}, which gives potential to optimise these systems for specific applications. \\

In this work, we apply experimental measurements and numerical modeling to study reactive species production in a gridded SDBD system, which is intended for application in chemical conversion processes. In such applications, the flow rate of gas through the system is an important control parameter, as it has a strong influence on the degree of conversion \cite{boeddecker2022, Mohsenimehr2023}. Therefore, the key focus of this work is on the influence of the gas flow rate on the mode of operation of the discharge. Experimentally, optical absorption spectroscopy is applied to measure reactive species densities. A semi-empirical numerical model is implemented to explain and better understand the experimentally measured density profiles. This model is based on those developed by Shimizu~\textit{et al.} \cite{shimizu2012} and Park~\textit{et al.} \cite{park2018}, but is extended to include two regions of differing species densities, based on the geometry of the SDBD system. The experimental setup is described in section~\ref{chap:experiment} and the numerical model is described in section~\ref{chap:modeling}. The results are described and discussed in section~\ref{chap:results}, and conclusions presented in section~\ref{chap:conclusion}.

\section{Experimental setup}
\label{chap:experiment}

\subsection{Surface dielectric barrier discharge}
\label{chap:sdbd}
\begin{figure}[ht]
    \centering
    \includegraphics[width=0.45\textwidth]{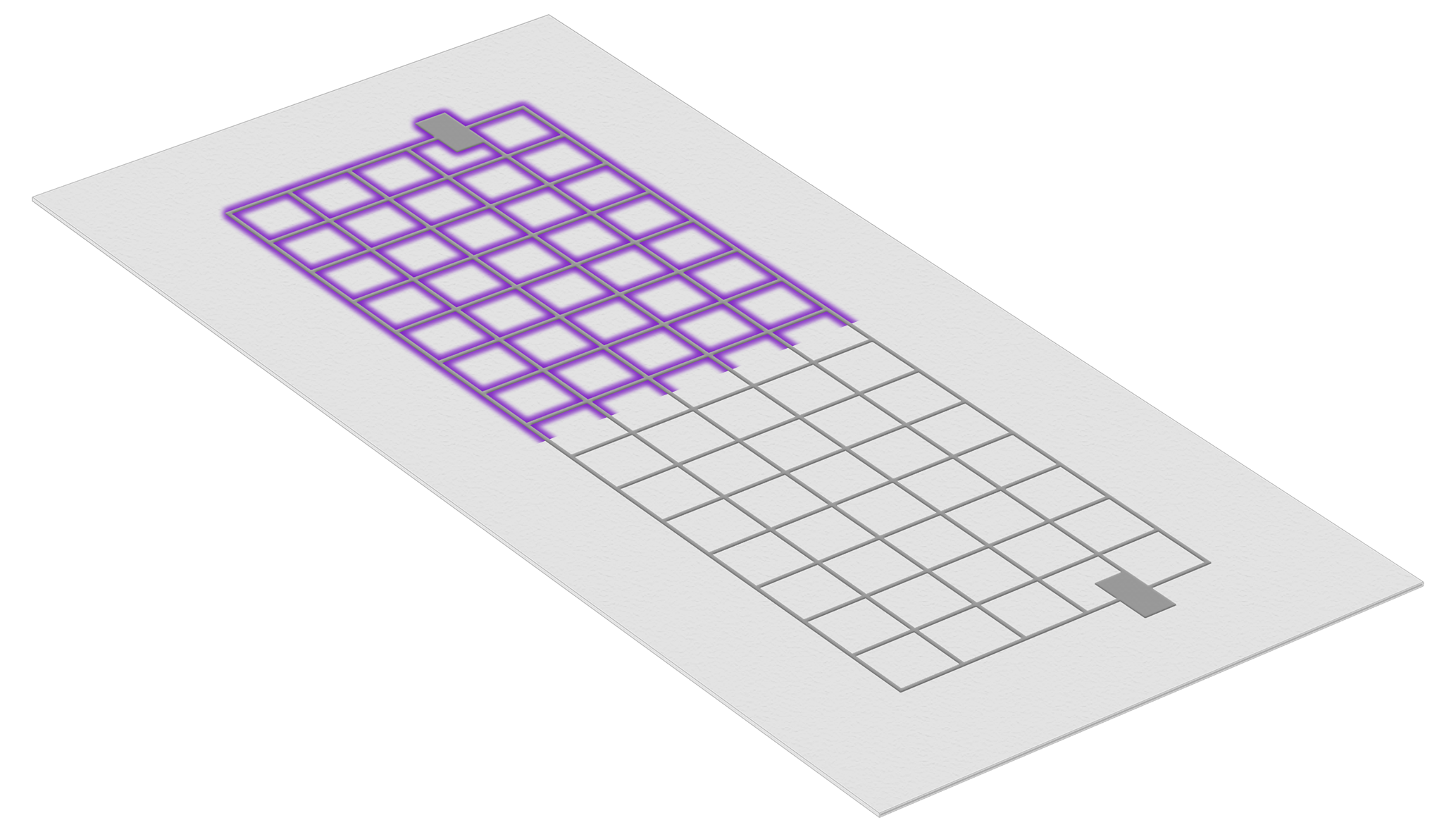}
	\caption{Sketch of the electrode configuration used to ignite the SDBD. The plasma ignites on both sides of the system, directly next to the driven (top) and grounded (bottom) metallic grid. For illustrative purposes the image is split into two parts, with (top half) and without (bottom half) an active plasma. \cite{schuecke2020} \textcopyright\,Institute of Physics (the “Institute”) and IOP Publishing Limited 2019.}
    \label{fig:electrode}
\end{figure}

The surface dielectric barrier discharge used for all measurements in this work has been studied in detail in previous works \cite{schuecke2020, schuecke2022, offerhaus2017, kogelheide2019}. The discharge is ignited on an $\alpha$-aluminum oxide plate with two symmetrical metallic grids printed onto both sides, which serve as driven and grounded electrode, respectively. They are separated by the aluminum oxide plate, which acts as the dielectric barrier, and has outer dimensions of 190\,x\,88\,x\,0.635\,mm. The metallic grid is 150\,x\,50\,mm in size, 0.45\,mm wide, and has a lattice constant of 10\,mm. The discharge is driven by voltages in the order of 8 to 13\,kV and ignites directly to the side of the metallic grid and on both sides of the dielectric surface. The plasma occupies a height of roughly 0.1\,mm and has a width of roughly 1\,mm \cite{offerhaus2018}. A sketch of the whole configuration is shown in figure~\ref{fig:electrode}. \\

The discharge is driven by a damped sine wave provided by a high voltage generator with external transformer (Redline G2000, Redline Technologies, Germany) at repetition frequencies of 250 to 4000\,Hz and a resonance frequency of about 86\,kHz. The damped sine wave is the result of a series resonant circuit formed by the transformer's inductance and the electrode's capacitance. The rectangular voltage pulses applied to the primary side of the transformer are converted into the kilovolt range on the secondary side, while exciting the resonance circuit. The plasma ignites several times per pulse, while the voltage of the damped sine wave still exceeds the ignition voltage. The system is capable of driving discharges with dissipated powers in the range of 0.5\,W at 250\,Hz and 8\,kV to 70\,W at 4\,kHz and 11\,kV. Further details are given in the work of Schücke~\textit{et al.} \cite{schuecke2020}. \\

For all measurements in this work dry synthetic air (ALPHAGAZ 1 Luft, AIR LIQUIDE Deutschland GmbH, Germany) at flow rates of 0\,slm to 10\,slm is applied using mass flow controllers (EL-FLOW Select, Bronkhorst High-Tech B.V., Netherlands). The corresponding gas stream velocity is in the range of 0.0\,ms\textsuperscript{-1} to 0.1\,ms\textsuperscript{-1}, at a constant pressure of 1\,bar. The shortest average residence time, along the active part of the electrode and for the maximum flow of 10\,slm, is around 1.5\,s.

\subsection{Reactor chamber and devices}
\label{chap:chamber}
\begin{figure}[ht]
    \centering
    \includegraphics[width=0.48\textwidth]{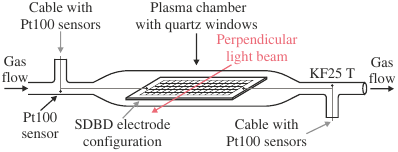}
	\caption{Schematic drawing of the reactor chamber and periphery. Optical absorption measurements are performed either across the electrode configuration or through the KF25 cross attached to the chamber outlet. The power connectors are not shown for the sake of simplicity. Adapted from \cite{schuecke2022} under CC BY 4.0.}
    \label{fig:setup}
\end{figure}

Like the discharge itself, the surrounding experimental setup and methodology used for the measurements in this work are mostly the same as presented by Schücke~\textit{et al.} in \cite{schuecke2022}. A reactor chamber made from aluminum, together with stainless steel vacuum parts, is used to ensure well defined operating conditions. In contrast to the chamber used in \cite{schuecke2020}, the one used in this work has been modified to provide larger areas of optical access through now three rectangular quartz windows (GVB GmbH, Germany). Two small windows, with dimensions of 75\,mm\,x\,15\,mm, are located to each side of the discharge along the gas flow direction, in order to allow an optical path across the reactor's width. A third, larger window is embedded into the lid of the reactor and allows optical access to the surface of the electrode configuration. Since all measurements in this work are performed in transmission, only the two smaller side windows are used here. A schematic drawing of the setup is shown in figure~\ref{fig:setup}. \\

The setup is fully automated using a modular PLC (BC9000, Beckhoff Automation GmbH \& Co. KG, Germany) and a digital user interface (LabVIEW 2019, National Instruments Corporation, USA), allowing for precise and reproducible measurements.

\section{Diagnostic methods}
\label{chap:diagnostics}

\subsection{Optical absorption spectroscopy}
\label{chap:oas}
Optical absorption spectroscopy is a technique used to determine absolute densities of specific species in an otherwise transparent medium. The technique is based on the Lambert-Beer law, according to equation~\ref{eq:lambert_beer}, which describes the difference in the intensity of light after being transmitted through the absorbing medium.
\begin{equation}
    \label{eq:lambert_beer}
        I_1 = I_0\cdot \exp(-\sigma( \lambda,T)\cdot n\cdot l)
\end{equation}

The properties of the medium are given by its density $n$, its effective absorption cross section $\sigma(\lambda,T)$ (where $\lambda$ is the wavelength and $T$ the temperature) and the absorption length $l$. The light that passes through the medium $I_0$ is partially absorbed, resulting in a lower intensity $I_1$ after transmittance. Under most circumstances ambient light and the light emitted by the plasma would have to be taken into consideration. However, in the presented case, the light emitted from the plasma does not reach the detector due to the plasma's very thin nature, and the ambient light was dimmed for all experiments. This means, that the Lambert-Beer law can be solved for the density, as shown in equation~\ref{eq:lambert_beer_2}.
\begin{equation}
    \label{eq:lambert_beer_2}
        n = \frac{-\ln\left(\frac{I_1}{I_0}\right)}{\sigma(\lambda,T)\cdot l}
\end{equation}

\begin{figure}[ht]
    \centering
    \includegraphics[width=0.48\textwidth]{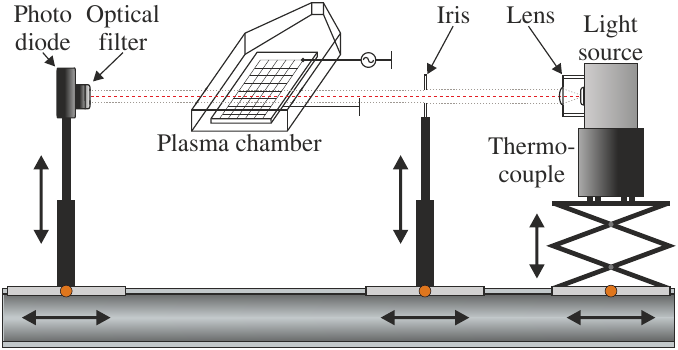}
	\caption{Schematic of the optical absorption spectroscopy setup. Adapted from \cite{schuecke2022} under CC BY 4.0.}
    \label{fig:oas}
\end{figure}

\begin{figure}[ht]
    \centering
    \includegraphics[width=0.48\textwidth]{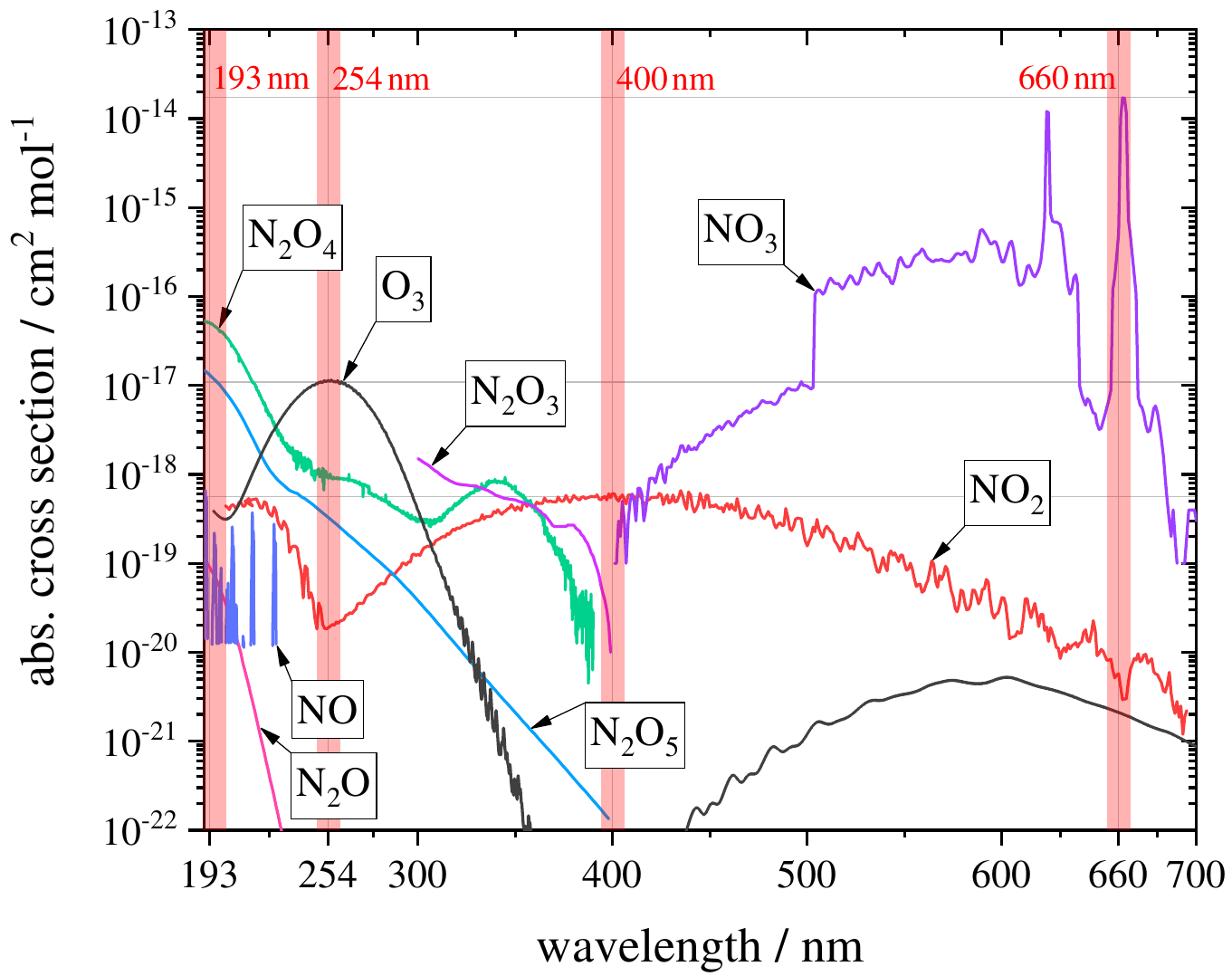}
	\caption{Absorption cross sections of the considered reactive oxygen and nitrogen species \cite{mpi2013}. Wavelength sections accepted by the band pass filters (see table~\ref{tab:filters}) are highlighted in red. Reproduced from \cite{schuecke2022} under CC BY 4.0.}
    \label{fig:abs_cross_sec}
\end{figure}

In this work specifically, the measured absolute densities of reactive oxygen and nitrogen species are acquired by a setup for OAS as shown in figure~\ref{fig:oas}. The method is the same as previously described in \cite{schuecke2022}. A thermally stabilized (TECMount 284, Arroyo Instruments LLC, United States) broadband light source (EQ-99X LDLS, Energetiq Technology Inc., United States) emits divergent light, that is then focused into a parallel beam by a plano-convex quartz collimating lens (\#48-274, Edmund Optics Ltd., United Kingdom) and restricted by an iris diaphragm. The latter eliminates possible distortions due to lens's edge effects, and also restricts reflections. After passing through the reactor chamber the light beam, which is oriented perpendicular to the gas flow, falls onto a photodiode (APD440A2, Thorlabs Inc., United States). The active area of the photodiode is circular and has a diameter of 1\,mm, which faces the reactor chamber's side window. It is positioned in the center of the light spot projected by the light source. The absorption length $l$ is given by the reactor's width of 10\,cm. The front of the photodiode is covered by an optical interference filter chosen specifically for the desired wavelength, based on the absorption cross section $\sigma(\lambda)$ for each species, as highlighted in figure~\ref{fig:abs_cross_sec}. \\

Two criteria are considered when selecting the wavelengths at which OAS is performed for each reactive species: the absorption cross section of each respective species in the selected interval should be as high as possible while also being sufficiently separated from those of the other species at that interval. A list of the bandpass filters (Edmund Optics Ltd., United Kingdom) that were chosen based on these criteria can be found in table~\ref{tab:filters}. \\

The difference in absorption cross section at 193\,nm, for N\textsubscript{2}O\textsubscript{4} and N\textsubscript{2}O\textsubscript{5}, is only about 72\,\%, which means that absorption caused by these two species can't be sufficiently separated. In our previous work, absorption at this wavelength was interpreted as a superposition of both species densities \cite{schuecke2022}. Here, we update this interpretation based on measurements in the literature, which indicate that densities of N\textsubscript{2}O\textsubscript{4} are significantly lower than those of N\textsubscript{2}O\textsubscript{5} in SDBDs, under broadly similar conditions to those in the system studied in this work \cite{huh2024}. Based on this, we interpret absorption at 193\,nm to be caused by N\textsubscript{2}O\textsubscript{5} and O\textsubscript{3}. Using the density of O\textsubscript{3}, known from the absorption at 254\,nm, the contribution of O\textsubscript{3} at 193\,nm can be calculated and the remaining absorption attributed to N\textsubscript{2}O\textsubscript{5}. In this way, the density of N\textsubscript{2}O\textsubscript{5} can be measured. In addition to N\textsubscript{2}O\textsubscript{5} and O\textsubscript{3}, NO\textsubscript{2} is also measured by absorption at 400\,nm. No significant amount of NO\textsubscript{3} could be detected, using the absorption line at 660\,nm. \\

\begin{table}[htbp]
    \centering
    \begin{tabular}{@{}lcc@{}}
    \toprule
    species & filter range\,/\,nm & EO item number \\ \midrule
    \ce{N2O4}, \ce{N2O5} & 193.0\,$\pm$\,7.5 & \#67-836 \\
    \ce{O3} & 254.0\,$\pm$\,5.0 & \#67-808 \\
    \ce{NO2} & 400.0\,$\pm$\,5.0 & \#65-677 \\
    \ce{NO3} & 660.0\,$\pm$\,5.0 & \#11-981 \\ \bottomrule
    \end{tabular}
    \caption{List of measured species with the respectively used filters' center wavelength and full width at half maximum, as well as the supplier's (Edmund Optics Ltd., United Kingdom) item number. Reproduced from \cite{schuecke2022} under CC BY 4.0.}
    \label{tab:filters}
\end{table}

\section{Modeling}
\label{chap:modeling}

\begin{figure}[ht]
   \centering
   \includegraphics[width=0.45\textwidth]{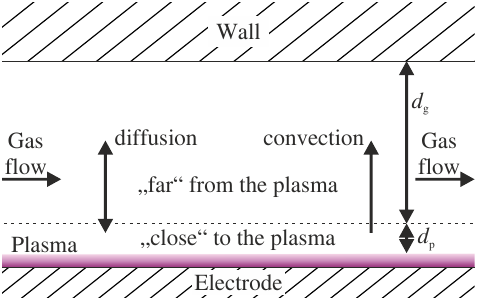}
   \caption{Schematic drawing of the spatial assumptions considered for transport within the model (not to scale). In the gas layer ``close" to the plasma constant in time densities of O, N\textsubscript{2}(A\textsuperscript{3}$\Sigma$), and O\textsubscript{2}(a\textsuperscript{1}$\Delta$), as well as a constant production rate of N are assumed.}
   \label{fig:schematic_model}
\end{figure}

A semi-empirical two-zone zero-dimensional chemistry model based on works by Sakiyama~\textit{et al.}, Shimizu~\textit{et al.}, and Park~\textit{et al.} is used to gain further insight into the reaction kinetics of oxygen and nitrogen species in the discharge and the surrounding neutral gas \cite{sakiyama2012, shimizu2012, park2018}. The model is divided into one zone ``close" to the plasma and one zone ``far" from the plasma, to allow for different species densities and gas temperatures in each region, as is expected to be the case in the experimental system. Reactive species that are formed in the ``close" to plasma volume are transported to the ``far" from plasma region due to diffusion and drift, caused by plasma-induced convection. A simplified schematic drawing of the geometric features and transport processes is given in figure~\ref{fig:schematic_model}. \\

As for the models of Shimizu~\textit{et al.} and Park \textit{et al.} \cite{shimizu2012, park2018}, the properties of the plasma are not simulated directly. Instead, the chemistry is mainly driven by assuming a fixed density of atomic oxygen (\textit{n}\textsubscript{O}), which drives the formation of O\textsubscript{3}, and a temporally varying density of highly vibrationally excited N\textsubscript{2}, with vibrational level $v>\mathrm{12}$ (\textit{n}\textsubscript{N\textsubscript{2}(\textit{v}\,$\geq$\,12)}). As described above, the combination of both species forms NO, which leads to a mode transition after a certain period of time, when \textit{n}\textsubscript{N\textsubscript{2}(\textit{v}\,$\geq$\,12)} reaches a sufficient level. In contrast to the models of Shimizu~\textit{et al.} and Park~\textit{et al.} \cite{shimizu2012, park2018}, the model developed here assumes atomic oxygen is only present in the zone ``close" to the plasma. This assumption is motivated by the fact that atomic oxygen is mainly produced in the plasma, and is sufficiently short lived that it cannot be transported over large distances away from the plasma. On the other hand, vibrationally excited N\textsubscript{2} molecules are long-lived and are expected to be distributed more evenly in the volume of the reactor, so the densities of these molecules are assumed the same in both regions. The temporal evolution of \textit{n}\textsubscript{N\textsubscript{2}(\textit{v}\,$\geq$\,12)} is given by:
\begin{equation}
    \label{eq:nN2v}
        n_{\mathrm{N_2}(v)} = n_{\mathrm{N_2}}F_{v>12} = n_{\mathrm{N_2}} \exp \left(-\frac{12 \Delta \varepsilon_v}{k_\mathrm{B} T_\mathrm{v}}\right),
\end{equation}
\begin{equation}
    \label{eq:Tv}
        T_{\mathrm{v}} = T_{\mathrm{g}} + T^{\mathrm{0}}_{\mathrm{v}}\left[1-\exp\left(\frac{-t}{\tau_v}\right)\right]
\end{equation}

In these equations, $\varepsilon_v$ represents the vibrational energy for a harmonic oscillator, k\textsubscript{B} is the Boltzmann constant, \textit{T}\rlap{\textsuperscript{0}}\textsubscript{v} describes the vibrational temperature at steady state, $\tau$\textsubscript{v} is the time constant over which the vibrational temperature increases, \textit{T}\textsubscript{g} is the gas temperature, and \textit{T}\textsubscript{v} is the vibrational temperature. As noted in Shimizu~\textit{et al.} \cite{shimizu2012}, the expression above implies a Maxwellian vibrational distribution function, which is unlikely to be the case in the experiment. As a result, the vibrational temperatures used here are not expected to be quantitatively equivalent to those in the experimental system, they are rather used to approximate the density of \textit{n}\textsubscript{N\textsubscript{2}(\textit{v}\,$\geq$\,12)}, which could also be achieved with different shapes of vibrational distribution function. \\

In addition to atomic oxygen and vibrationally excited nitrogen, fixed densities of N\textsubscript{2}(A\textsuperscript{3}$\Sigma$) and O\textsubscript{2}(a\textsuperscript{1}$\Delta$) are also assumed, following the approach of Park~\textit{et al.} \cite{park2018}. Lastly, a temporally fixed production term for atomic nitrogen, \textit{r}\textsubscript{N} is also included, as given in R17 of table~\ref{tab:reactions}. This is designed to approximate atomic nitrogen formation via reactions that do not rely on vibrationally excited nitrogen, and can occur both before and after the densities of vibrationally excited nitrogen have increased, such as electron impact dissociation of N\textsubscript{2}. This is implemented as a fixed production rate, rather than a fixed density of N, so that the variation of N density as a result of the temporally varying rate of R9 (see table~\ref{tab:reactions}) can also be accounted for. \\

Overall, the chemical model has eight free parameters: \textit{n}\textsubscript{O}, \textit{T}\rlap{\textsuperscript{0}}\textsubscript{v}, $\tau$\textsubscript{v}, \textit{T}\textsubscript{g}, \textit{T}\textsubscript{p}, \textit{n}\textsubscript{N\textsubscript{2}(A\textsuperscript{3}$\Sigma$)}, \textit{n}\textsubscript{O\textsubscript{2}(a\textsuperscript{1}$\Delta$)}, and \textit{r}\textsubscript{N}. These parameters drive the overall chemistry, initiated by atomic oxygen, atomic nitrogen, N\textsubscript{2}(A\textsuperscript{3}$\Sigma$), O\textsubscript{2}(a\textsuperscript{1}$\Delta$), and vibrationally excited nitrogen. As is described later, these eight free parameters are empirically varied in order to fit the simulated species densities to those measured in the experiment. \\

In contrast to reality, a simplified, homogeneous plasma layer that generates a constant in time density of O, N\textsubscript{2}(A\textsuperscript{3}$\Sigma$) and O\textsubscript{2}(a\textsuperscript{1}$\Delta$), and production rate of N is assumed. The pulsed nature of the discharge is neglected, which may lead to significant differences between measurements and model, as will be further illustrated in a later section. Overall, the O density, N\textsubscript{2}(A\textsuperscript{3}$\Sigma$) density, O\textsubscript{2}(a\textsuperscript{1}$\Delta$) density, and the production rate of N in the model are not necessarily reflective of the instantaneous values of these quantities, but are rather effective values that average over spatially and temporally variable profiles. \\

An expression for the transport of species between the two regions of the model by diffusion, considering the difference in height between ``close" to plasma volume and ``far" from plasma volume, is adapted from Sakiyama~\textit{et al.} \cite{sakiyama2012}, as shown in equation~\ref{eq:diffusion}, using the diffusion coefficients D given in table~\ref{tab:diffusion}. \\

\begin{table}[ht]
    \centering
    \begin{tabular}{@{}ll@{}}
    \toprule
    Species\; & D\,(m\textsuperscript{2}s\textsuperscript{-1}) \\ \midrule
    \ce{O3}   & $1.5 \times 10^{-5}$ \\
    \ce{N}    & $2.9 \times 10^{-5}$ \\
    \ce{NO}   & $2.0 \times 10^{-5}$ \\
    \ce{NO2}  & $1.7 \times 10^{-5}$ \\
    \ce{NO3}  & $0.9 \times 10^{-5}$ \\
    \ce{N2O}  & $1.6 \times 10^{-5}$ \\
    \ce{N2O3} & $1.0 \times 10^{-5}$ \\
    \ce{N2O4} & $1.0 \times 10^{-5}$ \\
    \ce{N2O5} & $1.0 \times 10^{-5}$ \\
    \bottomrule
    \end{tabular}
    \caption{Diffusion coefficients for the considered reactive species as derived by Sakiyama~\textit{et al.} from Bird~\textit{et al.}, using gas kinetic theory \cite{sakiyama2012, transport_phenomena}.}
    \label{tab:diffusion}
\end{table}

In the presented geometry the height of the visible plasma has been determined by Offerhaus~\textit{et al.} to be roughly 100\,µm. Here, the height of the region ``close" to the plasma \emph{d}\textsubscript{p} has been chosen as 1\,mm, based on the estimated diffusion length, accounting for chemical reactions, of atomic oxygen under the given conditions. The height of the surrounding gas stream up to the wall of the reactor \emph{d}\textsubscript{g} is fixed by the reactor geometry at 10\,mm. The densities of each reactive species in the plasma volume and the gas volume are given by \emph{n}\textsubscript{p} and \emph{n}\textsubscript{g}, respectively. Species transport between these regions is given by flux terms that relate to transport due to diffusion and plasma-induced convection. The flux due to diffusion is given by the following expression:
\begin{equation}
    \label{eq:diffusion}
        \Gamma_{\mathrm{diff}} = \frac{D(n_\mathrm{p}-n_\mathrm{g})}{\frac{1}{2}(d_\mathrm{p}+d_\mathrm{g})}
\end{equation}

The rate of transport due to drift is estimated based on Böddecker~\textit{et al.}~\cite{boeddecker2023, boeddecker2024}, who performed particle image velocimetry measurements (PIV) of an identical DBD system. In these works typical mean vertical gas velocities v\textsubscript{drift,pg} are in the order of 0.1\,ms\textsuperscript{-1}. These velocities are used to define the flux due to convection in the model according to equation~\ref{eq:drift}.
\begin{equation}
    \label{eq:drift}
        \Gamma_{\mathrm{drift}} = v_{\mathrm{drift}}\cdot n_\mathrm{p}
\end{equation}

As a third transport mechanism, a continuous gas flow, as supplied during the measurements, is considered. As the neutral gas density according to the ideal gas law is kept constant, and consists of 20.8\,\% oxygen and 79.2\,\% nitrogen, the applied gas flow removes all simulated species in accordance with equation~\ref{eq:k_flow}. Here, $\Phi_{gas}$ denotes the applied gas flow in standard liters per minute (slm) and 0.06 is a factor used to convert from slm to SI units. The gas particles are replaced as fractions of the total gas density, which cancels out with the gas density introduced from converting slm to rate, which is why it doesn't appear in equation~\ref{eq:k_flow}.
\begin{equation}
    \label{eq:k_flow}
        k_{\mathrm{flow;p,g}} = \frac{\Phi_{\mathrm{gas}}\mathrm{[slm]}\cdot n_{\mathrm{tot}}}{60\cdot1000\cdot V_{\mathrm{reactor}}}\cdot\frac{d_{\mathrm{p,g}}}{d_\mathrm{p} + d_\mathrm{g}}
\end{equation}

Finally, the total change of the density of each considered species is given by rate equations below supplemented by the individual transport processes as outlined above. These equations are solved for the species listed in table \ref{tab:diffusion}.
\begin{equation}
    \label{eq:dnp}
        \frac{dn_\mathrm{p}}{dt}=\sum_{j}k_j\prod n_{\mathrm{r},j}-\frac{\Gamma_{\mathrm{diff}}}{d_\mathrm{p}}-\frac{\Gamma_{\mathrm{drift}}}{d_\mathrm{p}}-k_{\mathrm{flow;p,g}}\cdot\frac{n_\mathrm{p}}{n_{\mathrm{tot}}}
\end{equation}
\begin{equation}
    \label{eq:dng}
        \frac{dn_\mathrm{g}}{dt}=\sum_{j}k_j\prod n_{\mathrm{r},j}+\frac{\Gamma_{\mathrm{diff}}}{d_\mathrm{g}}+\frac{\Gamma_{\mathrm{drift}}}{d_\mathrm{g}}-k_{\mathrm{flow;p,g}}\cdot\frac{n_\mathrm{g}}{n_{\mathrm{tot}}}
\end{equation}

\section{Results and discussion}
\label{chap:results}

\subsection{Measured densities of reactive species}
\label{chap:densities_measured}
\begin{figure}[ht]
   \centering
   \includegraphics[width=0.45\textwidth]{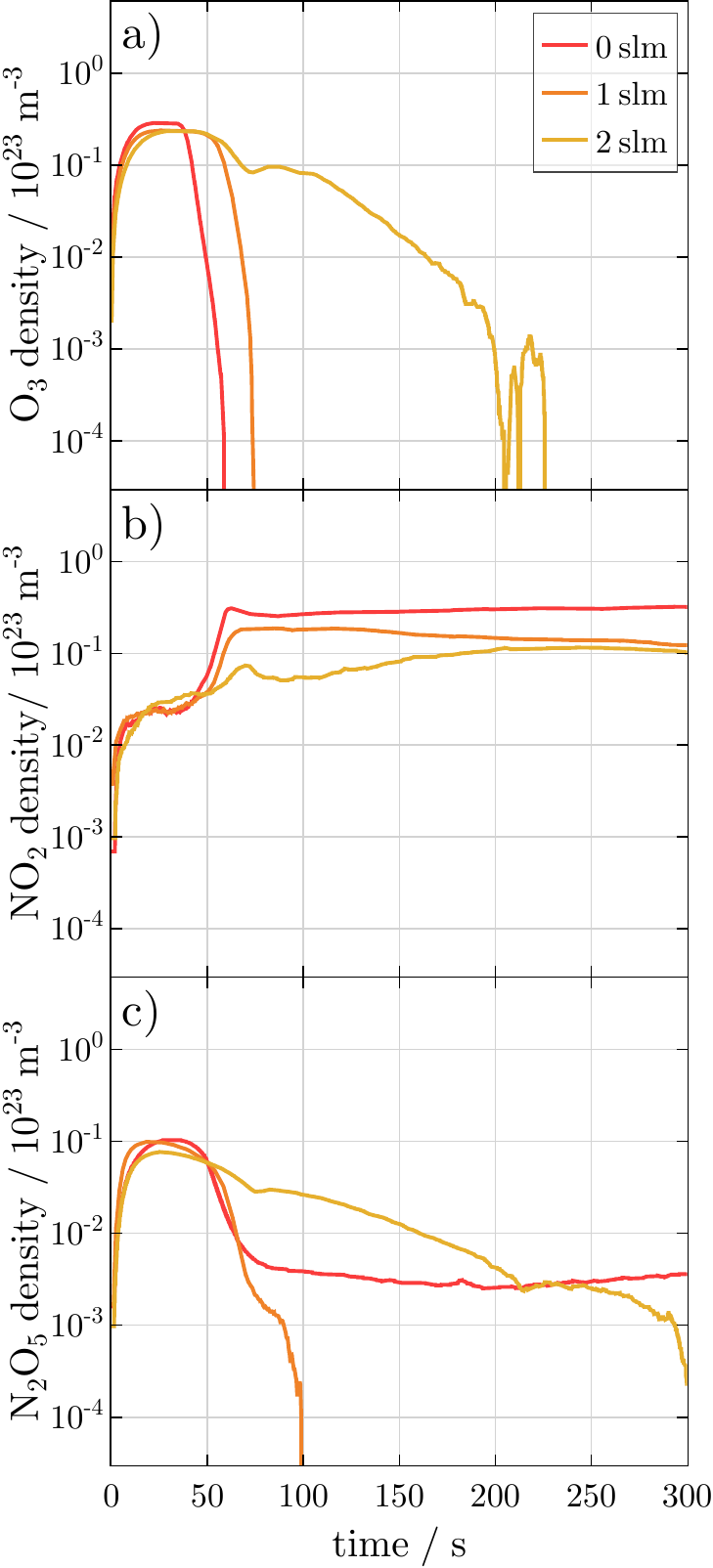}
   \caption{Measured densities of a)~O\textsubscript{3}, b)~NO\textsubscript{2}, and c)~N\textsubscript{2}O\textsubscript{5} during SDBD operation for gas flows of 0\,slm, 1\,slm and 2\,slm of dry synthetic air.}
   \label{fig:densities_measured012}
\end{figure}

Measured densities of the reactive species O\textsubscript{3}, NO\textsubscript{2}, and N\textsubscript{2}O\textsubscript{5} were obtained during SDBD operation, and for varying flow rates from 0\,slm to 10\,slm. Figure~\ref{fig:densities_measured012} illustrates the obtained densities for the species O\textsubscript{3}, NO\textsubscript{2}, and N\textsubscript{2}O\textsubscript{5}, for the lower flow rates of 0\,slm, 1\,slm, and 2\,slm, which are equivalent to flow speeds of 0\,ms\textsuperscript{-1}, 0.01\,ms\textsuperscript{-1}, and 0.02\,ms\textsuperscript{-1}, respectively. During the initial operation of the discharge, from 0\,s to approximately 40\,s, the densities for all three species increase significantly, approaching an intermediate steady state. After this initial increase, however, the densities of O\textsubscript{3} and N\textsubscript{2}O\textsubscript{5} decrease again, while the density of NO\textsubscript{2} increases again and even further. As discussed in the introduction, this behavior, which can be described as a distinct mode-transition, is known in literature and has been demonstrated and studied in other works, such as Shimizu~\textit{et al.}, Pavlivich~\textit{et al.}, or Park~\textit{et al.} \cite{shimizu2012, pavlovich2014, park2021}. These two modes are typically called the O\textsubscript{3} mode and the NO\textsubscript{x} mode, presenting a transition over operating time. \\

These measured results and the observed mode-transition can be analyzed in greater detail with the help of the reaction scheme given in table~\ref{tab:reactions}, which is also used for the simulations in the following section. For all cases, the discharge starts in the O\textsubscript{3} mode, which is driven by the electron impact dissociation of O\textsubscript{2}, into O, and the subsequent production of O\textsubscript{3}, according to R2 of table~\ref{tab:reactions}:
\begin{equation}
    \ce{O + O2 + M -> O3 + M}
\end{equation}

The mode transition is then initiated by the increasing vibrational temperature of molecular nitrogen, here simplified as N\textsubscript{2}(\textit{v}), when the highly vibrationally excited molecules ($v>12$) reach sufficient densities to dominate the chemical kinetics via reaction R9. This reaction describes the dissociation of N\textsubscript{2}, forming the products O and NO, the latter of which reacts further to produce NO\textsubscript{2} and N\textsubscript{2}O\textsubscript{5}. While the general behavior of the species densities is similar for flow rates between 0\,slm and 2\,slm, the time point at which the mode transition occurs later as the flow rate increases. This potentially implies a change in the densities of vibrationally excited \ce{N2}, as will be discussed in more detail in the next section. \\

\begin{figure}[ht]
   \centering
   \includegraphics[width=0.45\textwidth]{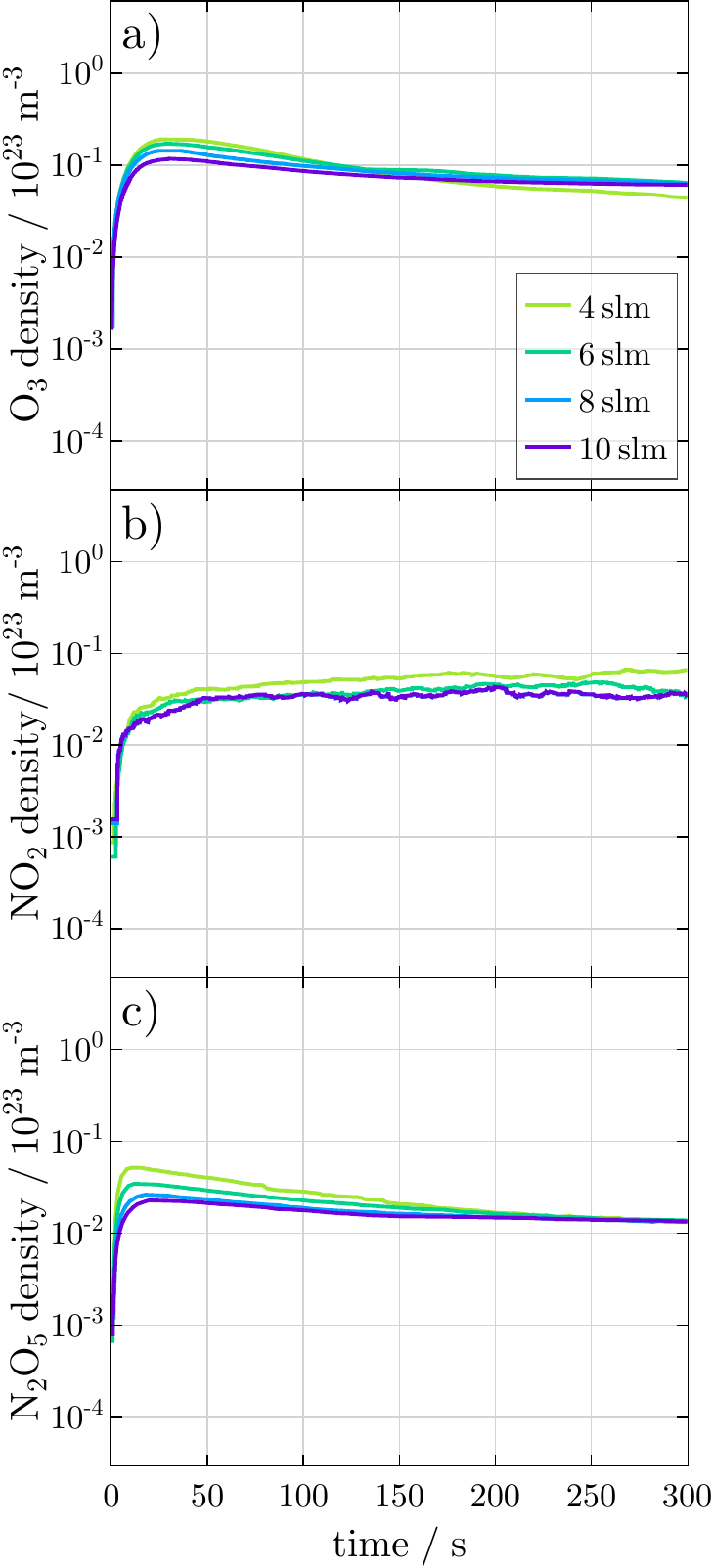}
   \caption{Measured densities of a)~O\textsubscript{3}, b)~NO\textsubscript{2}, and c)~N\textsubscript{2}O\textsubscript{5} during SDBD operation for gas flows of 4\,slm, 6\,slm, 8\,slm, and 10\,slm of dry synthetic air.}
   \label{fig:densities_measured46810}
\end{figure}

On the other hand, figure~\ref{fig:densities_measured46810} illustrates the temporal evolution of the densities measured at higher flow rates. In these cases, the distinct mode transition cannot be observed during the measurement. The behavior of the densities for the three species has the same initial increase as described before, and slowly approaches a steady state over several minutes, with the density of \ce{NO2} gradually increasing, and those of \ce{O3} and \ce{N2O5} gradually decreasing after the initial increase over roughly 10\,s. These differences, compared to the measurements at lower flow rates, can potentially be explained by the increase in the gas velocity, which reaches up to 0.1\,m\textsuperscript{-1} for a flow rate of 10\,slm, with increasing flow rate. A higher gas velocity reduces the residence time of the species in the chamber, thereby limiting interactions between species and hindering reactions that take place on longer timescales. Since the timescales required for the production of highly vibrationally excited \ce{N2} are comparatively long, it would be expected that their densities would decrease with decreasing residence time. This is broadly consistent with the lack of an observable mode transition at higher flow rates, as a lower level of vibrational excitation would lead to less NO production via reaction R9 and correspondingly less \ce{O3} consumption and lower densities of \ce{NO2}. These concepts will be elaborated on further in the next section, which focuses on the results of the numerical model.

\subsection{Comparison of experiment and model}
\label{chap:densities_modeled}
To better understand the interactions between species involved in the discharge, a two-zone zero-dimensional chemistry model was developed, as described in section~\ref{chap:modeling}. The model simulates the densities of the species involved in the reactions listed in table~\ref{tab:reactions}, for flow rates ranging from 0\,slm to 10\,slm. The objective of the model is to describe the chemical reactions that primarily drive the production and consumption of ozone and nitrogen oxides. \\

Initially, the model was developed based on the chemistry considered in previous works \cite{shimizu2012, park2018}, with several modifications. The chemical reactions presented in the table~\ref{tab:reactions} rely on macroscopic fitting parameters as drivers of the overall chemical system, as mentioned in section~\ref{chap:modeling}. These define the densities of O, vibrationally excited nitrogen, N\textsubscript{2}(A\textsuperscript{3}$\Sigma$), O\textsubscript{2}(a\textsuperscript{1}$\Delta$), the production rate of N and the rate of reactions that depend on the neutral gas temperature. The final values of these parameters are provided in table~\ref{tab:parameters}. To determine these values, initial estimates were taken from previous works for \textit{n}\textsubscript{O}, \textit{T}\rlap{\textsuperscript{0}}\textsubscript{v}, and $\tau$\textsubscript{v}, \textit{n}\textsubscript{N\textsubscript{2}(A\textsuperscript{3}$\Sigma$)}, \textit{n}\textsubscript{O\textsubscript{2}(a\textsuperscript{1}$\Delta$)} and room temperature was assumed for the gas temperatures in both regions \textit{T}\textsubscript{g} and \textit{T}\textsubscript{p}. Subsequently, the influence of various parameters was studied by systematically changing them individually. Once the impact of each parameter was understood, the model was compared with measurements to adjust the necessary parameters for a flow rate of 0\,slm. Finally, by running simulations where these parameters are varied at each flow rate, and comparing the results with the experimental measurements, the final values of each parameter, for the considered flow conditions, were determined. It should be emphasised that the parameters' combinations are not necessarily unique in giving a certain level of agreement with the experimental data, and that a different combination of these parameters may also lead to good agreement with the experimental data. Nevertheless, the general trends in these parameters can provide some insight into the reasons for the changing temporal profiles as a function of the gas flow rate. \\

As described in the previous sections, the density of highly vibrationally excited nitrogen directly influences the reaction dynamics, and is described by equations~\ref{eq:nN2v} and~\ref{eq:Tv}, where the parameters listed in table~\ref{tab:parameters} define the density \textit{n}\textsubscript{N\textsubscript{2}($v>12$)} as a function of time. These parameters include \textit{T}\textsubscript{p}, which is the gas temperature in the ``close" to the plasma region, \textit{T}\textsubscript{g}, which is the gas temperature in the ``far" from plasma region, and \textit{T}\rlap{\textsuperscript{0}}\textsubscript{v} and $\tau$\textsubscript{v}, which define the temporal evolution of \textit{n}\textsubscript{N\textsubscript{2}($v>12$)}. In the following, together with \textit{n}\textsubscript{O}, these values are referred to as fitting parameters, and their influence on the simulated densities is discussed to better understand the mechanisms of NO generation from N\textsubscript{2}($v>12$), according to reaction R9 in table~\ref{tab:reactions}, and its role in the suppression of O\textsubscript{3} formation through reaction R18. As mentioned above, in order to get insight on the qualitative effects of each of the fitting parameters on the simulated density profiles, a manual variation of the parameters has been performed and their influence will be explained in the following. \\

\begin{itemize}
    \item An increase of \textit{T}\textsubscript{g} leads to an earlier transition between the O\textsubscript{3} and NO\textsubscript{x} modes, which is likely related to the gas temperature dependencies of O\textsubscript{3} production and consumption. The rate coefficient for O\textsubscript{3} production via reaction R2 decreases with increasing gas temperature, while the rate coefficients for several O\textsubscript{3} consumption reactions R18 to R21 increase with gas temperature. The combined effect is an earlier transition to the NO\textsubscript{x} mode. \\
        
    \item An increase in the steady-state vibrational temperature \textit{T}\rlap{\textsuperscript{0}}\textsubscript{v}, also leads to an earlier transition between the O\textsubscript{3} and NO\textsubscript{x} modes, because \textit{T}\rlap{\textsuperscript{0}}\textsubscript{v} defines the temporal variation of the vibrational temperature, and therefore the N\textsubscript{2}($v>12$) density. With more N\textsubscript{2}($v>12$) molecules, their reaction with atomic oxygen, as described in reaction R9 in table~\ref{tab:reactions}, becomes more frequent. Consequently, more NO is produced. This increases the rate of O\textsubscript{3} consumption by R18. The result is a shorter duration of the O\textsubscript{3} mode, leading to lower maximum O\textsubscript{3} densities and an earlier transition to the NO\textsubscript{x} mode. \\

    \item A change of the vibrational time constant $\tau$\textsubscript{v}, affects the expression that determines the onset time of the increasing vibrational temperature \textit{T}\textsubscript{v}. Therefore, changing the value of $\tau$\textsubscript{v} causes a shift of the densities over time, i.e., the rapid initial production of NO starts later with larger values of $\tau$\textsubscript{v} and, in the same way, the transition time between the modes also occurs later. \\

    \item An increase in the value of \textit{r}\textsubscript{N} tends to lead to higher rates of NO production via reactions R14 to R16, leading to an earlier increase in the density of NO\textsubscript{2}. However, large values of \textit{r}\textsubscript{N} tend to decrease the steady state densities of NO\textsubscript{2} and N\textsubscript{2}O\textsubscript{5}, as N atoms also contribute to consumption of NO and NO\textsubscript{2} via reactions R11 and R12. \\

    \item By increasing the densities of N\textsubscript{2}(A\textsuperscript{3}$\Sigma$) and O\textsubscript{2}(a\textsuperscript{1}$\Delta$), it was found that the mode transition occurs a few seconds earlier, the maximum value of O\textsubscript{3} decreases, and is reached earlier. Although both species produce this effect, the impact of O\textsubscript{2}(a\textsuperscript{1}$\Delta$) is lower. The influence of these densities can be understood through reactions R22 and R23 for the case of N\textsubscript{2}(A\textsuperscript{3}$\Sigma$), and R21 for the case of O\textsubscript{2}(a\textsuperscript{1}$\Delta$). These reactions lead to higher O\textsubscript{3} consumption, explaining the decrease in its maximum density. On the other hand, increasing these species' densities allows for a faster increase in the densities of NO\textsubscript{2} and N\textsubscript{2}O\textsubscript{5}, as reactions such as R8 and R23 promote greater NO production.
\end{itemize}

\begin{figure}[ht]
   \centering
   \includegraphics[width=0.45\textwidth]{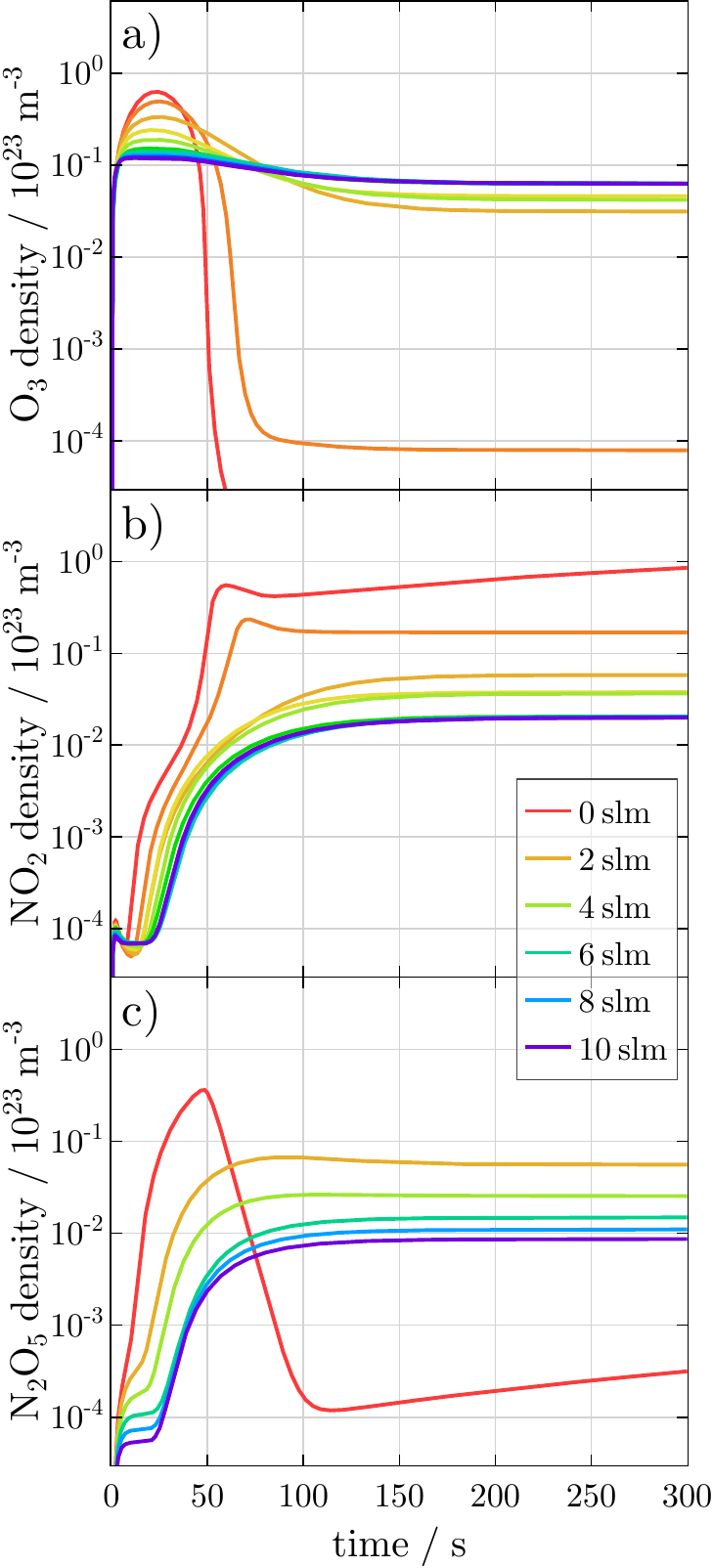}
   \caption{Simulated densities of a)~O\textsubscript{3}, b)~NO\textsubscript{2}, and c)~N\textsubscript{2}O\textsubscript{5} during SDBD operation for gas flows of 0\,slm, 2\,slm, 4\,slm, 6\,slm, 8\,slm, and 10\,slm of dry synthetic air.}
   \label{fig:densities_simulated}
\end{figure}

\begin{figure}[ht]
   \centering
   \includegraphics[width=0.45\textwidth]{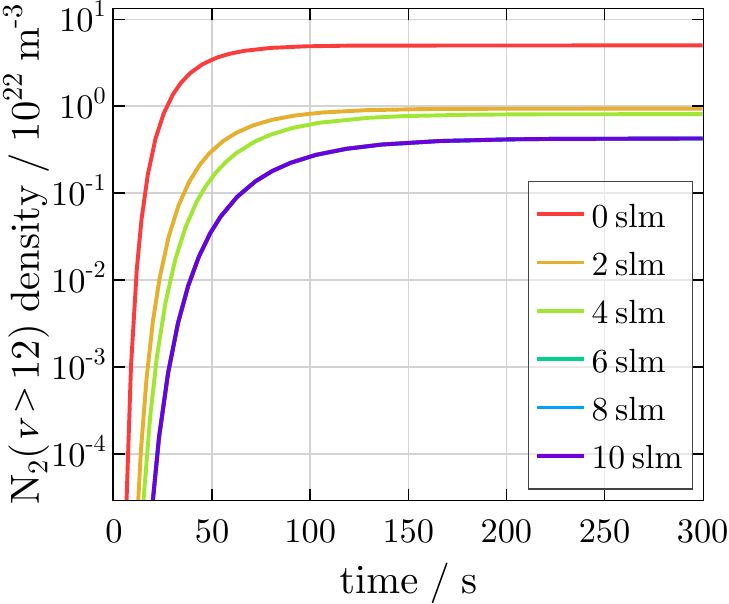}
   \caption{Exemplary simulated density profiles of N\textsubscript{2}($v>12$) for the parameter set given in table~\ref{tab:parameters}.}
   \label{fig:density_nN2v}
\end{figure}

The results obtained by the simulation can now be analyzed and compared to the experimental results, for more detailed insight. Figure~\ref{fig:densities_simulated} illustrates the simulated densities for O\textsubscript{3}, NO\textsubscript{2} and N\textsubscript{2}O\textsubscript{5} with flow rate increments of 2\,slm. Once again, the mode transition from the O\textsubscript{3} mode to the NO\textsubscript{x} mode is observed at low flow rates, while at higher flow rates, a steady state is achieved after a transition period. To analyze the influence of flow rate on the density of \textit{n}\textsubscript{N\textsubscript{2}($v>12$)}, as  well as the influence on the final densities obtained for the other species, figure \ref{fig:density_nN2v} presents the density over time for varying flow rates. \\

A comparison of figures \ref{fig:densities_simulated} and \ref{fig:density_nN2v} shows that an increase in gas flow rate leads to a decrease in \textit{n}\textsubscript{N\textsubscript{2}($v>12$)}. This supports the hypothesis presented in the previous chapter, that this variation is a key factor for the absence of the mode transition at higher flow rates, as the decrease in \textit{n}\textsubscript{N\textsubscript{2}($v>12$)} is evident. \\

\renewcommand{\arraystretch}{1.5}
\begin{table*}[ht]
    \centering
    \begin{tabular}{@{}rlccccccccccc@{}}
        \toprule
        \multicolumn{2}{c}{\multirow{2}{*}{Parameter}} & \multicolumn{11}{c}{Gas flow rate} \\
         & & 0\,slm & 1\,slm & 2\,slm & 3\,slm & 4\,slm & 5\,slm & 6\,slm & 7\,slm & 8\,slm & 9\,slm & 10\,slm \\
        \midrule
        $\mathrm{n_O}\cdot 10^{17}$ & $\mathrm{[m^{-3}]}$ & $7.30$ & $8.40$ & $8.50$ & $8.60$ & $8.70$ & $9.00$ & $10.00$ & $11.00$ & $12.00$ & $13.00$ & $14.00$ \\
        $\mathrm{O_2(a^1\Delta)}\cdot 10^{18}$ & $\mathrm{[m^{-3}]}$ & $1.00$ & $1.00$ & $1.00$ & $1.00$ & $1.00$ & $1.00$ & $1.00$ & $1.00$ & $1.00$ & $1.00$ & $1.00$ \\
        $\mathrm{N_2(A^3\Sigma)}\cdot 10^{15}$ & $\mathrm{[m^{-3}]}$ & $1.00$ & $1.00$ & $1.00$ & $1.00$ & $1.00$ & $1.00$ & $1.00$ & $1.00$ & $1.00$ & $1.00$ & $1.00$ \\
        $r_{\mathrm{N}}\cdot 10^{18}$ & $\mathrm{[m^{-3}s^{-1}]}$ & $1.00$ & $1.00$ & $1.00$ & $1.00$ & $1.00$ & $1.00$ & $1.00$ & $1.00$ & $1.00$ & $1.00$ & $1.00$ \\
        $T_{\mathrm{v}}^0$ & $\mathrm{[K]}$ & $6500$ & $6000$ & $5000$ & $4900$ & $4900$ & $4500$ & $4500$ & $4500$ & $4500$ & $4500$ & $4500$ \\
        $\tau_{\mathrm{v}}$ & $\mathrm{[s]}$ & $18$ & $25$ & $25$ & $25$ & $30$ & $30$ & $35$ & $35$ & $35$ & $35$ & $35$ \\
        $T_{\mathrm{g}}$ & $\mathrm{[K]}$ & $310$ & $310$ & $310$ & $310$ & $310$ & $310$ & $310$ & $310$ & $310$ & $310$ & $310$ \\
        $T_{\mathrm{p}}$ & $\mathrm{[K]}$ & $335$ & $335$ & $335$ & $335$ & $335$ & $340$ & $340$ & $340$ & $340$ & $340$ & $340$ \\
        \bottomrule
    \end{tabular}
  \caption{Parameters used to fit the simulated densities to the measured ones, for each of the tested gas flow rates.}
  \label{tab:parameters}
\end{table*}

\begin{figure}[ht]
    \vspace{0.16cm}
    \centering
    \includegraphics[width=0.45\textwidth]{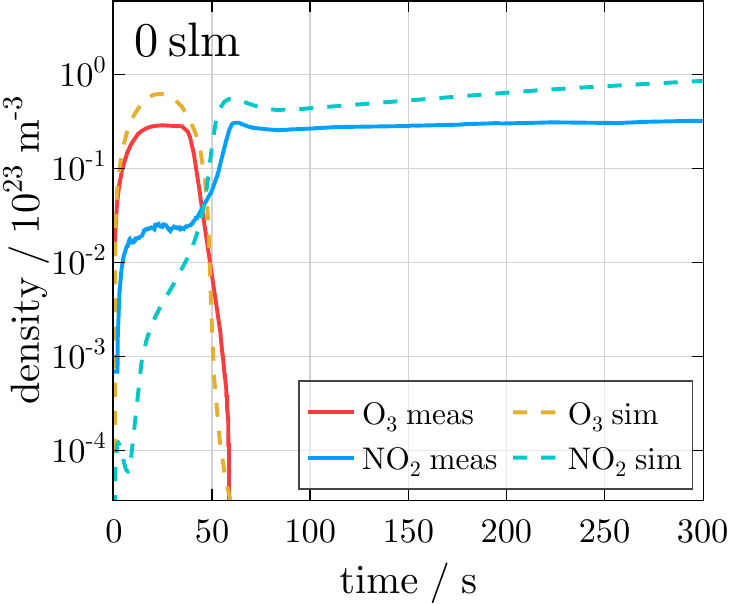}
    \caption{Comparison of a)~O\textsubscript{3}, b)~NO\textsubscript{2}, and c)~N\textsubscript{2}O\textsubscript{5} densities for measurement and simulation at a gas flow of 0\,slm of dry synthetic air.}
    \label{fig:0slm_comparison}
\end{figure}

Figure~\ref{fig:0slm_comparison} compares the measured and simulated densities of O\textsubscript{3} and NO\textsubscript{2} for a flow rate of 0\,slm. The results demonstrate a good qualitative agreement between the simulation and the measurements, with absolute agreement in the order of 50\,\%. Although the maximum and steady state density values are slightly higher in the simulation, they remain comparably close, when considering the full dynamic range of the densities. The experimentally measured behavior of O\textsubscript{3} is followed by the modeled density, with the temporal evolution matching the rise, peak, and decline observed in the measurements. For NO\textsubscript{2}, there is a slight discrepancy in the initial increase timing, however, the simulation aligns well with the time of the mode transition. \\

\begin{figure}[ht]
   \centering
   \includegraphics[width=0.45\textwidth]{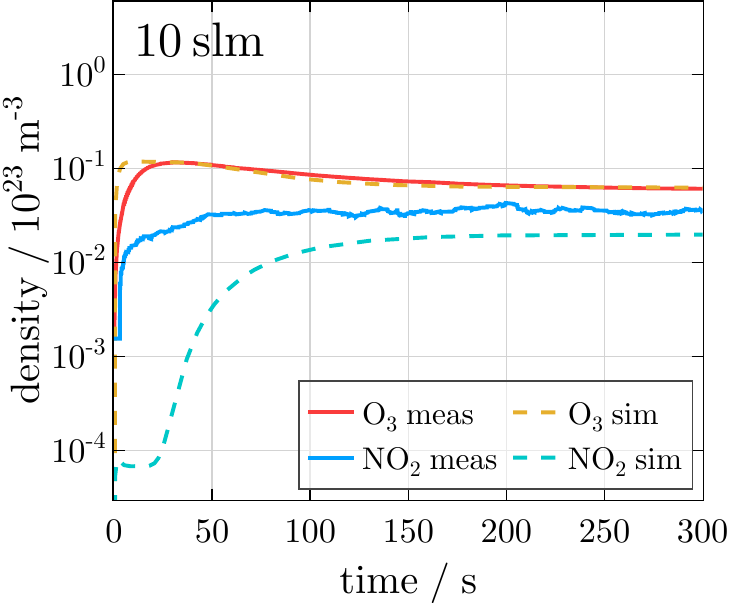}
   \caption{Comparison of simulated and measured densities of a)~O\textsubscript{3}, b)~NO\textsubscript{2}, and c)~N\textsubscript{2}O\textsubscript{5} for a flow rate of 10\,slm of dry synthetic air.}
   \label{fig:10slm_comparison}
\end{figure}

To evaluate the model’s response at higher flow rates, figure~\ref{fig:10slm_comparison} compares the measured and simulated densities at a flow rate of 10\,slm. The O\textsubscript{3} production is accurately represented in the model. However, NO\textsubscript{2} shows greater discrepancies as the flow rate increases, with the primary difference being a delayed onset of modeled NO\textsubscript{2} production compared with that observed in the experiment. One pathway for the production of NO\textsubscript{2} involves N to enable reactions R13 to R16 (see table~\ref{tab:reactions}). Comparing with figure~\ref{fig:density_nN2v}, where the density of N\textsubscript{2}($v>12$) at 10\,slm reaches a level close to steady state after 50\,s, in figure~\ref{fig:10slm_comparison} it is evident that simulated NO\textsubscript{2} only increases when N\textsubscript{2}($v>12$) archives sufficient density. One possibility for the delayed production of NO\textsubscript{2} at these higher flow rates may be that the model underestimates the production of N before the increase of N\textsubscript{2}($v>12$), hindering NO\textsubscript{2} formation at earlier time points. For these conditions, increasing the constant N atom source, reaction R17, in the model does lead to an earlier production of NO\textsubscript{2}, however, larger values also decrease the steady state NO\textsubscript{2} density. This points to the potential importance of N atoms for the accurate simulation of NO\textsubscript{2} densities, particularly at higher flow rates where the role of vibrationally excited nitrogen appears to be decreased. \\

\begin{figure}[ht]
   \centering
   \includegraphics[width=0.45\textwidth]{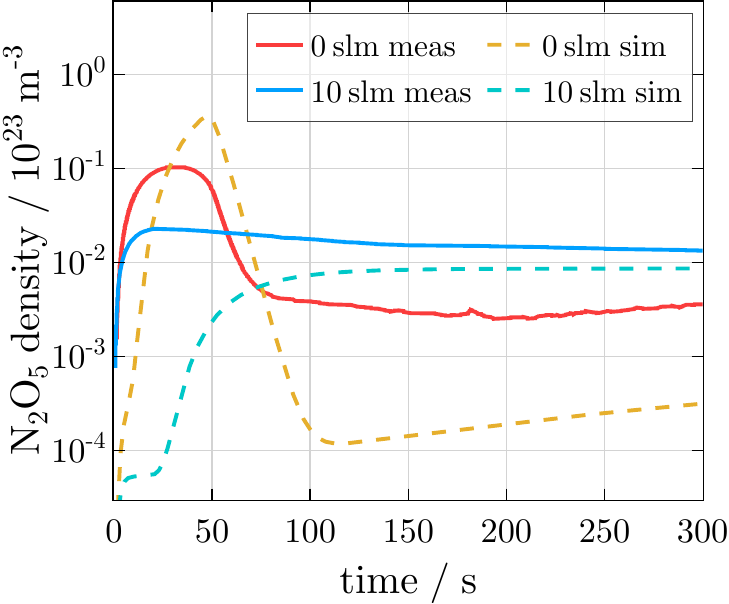}
   \caption{Comparison of simulated and measured densities of N\textsubscript{2}O\textsubscript{5} for gas flows of 0\,slm and 10\,slm of dry synthetic air.}
   \label{fig:N2O5_comparison}
\end{figure}

The production of N\textsubscript{2}O\textsubscript{5}, as described by reaction R28 in table~\ref{tab:reactions}, is determined by the densities of NO\textsubscript{2} and NO\textsubscript{3}. Therefore, analyzing the N\textsubscript{2}O\textsubscript{5} density at flow rates of 0\,slm and 10\,slm is beneficial for understanding the impact of NO\textsubscript{2} and, consequently, of NO on the broader chemistry. Figure~\ref{fig:N2O5_comparison} compares the measured and simulated N\textsubscript{2}O\textsubscript{5} densities at these flow rates. \\

The behavior observed at 10\,slm is consistent with the analysis for NO\textsubscript{2}. The main difference between measurements and model is the delayed onset of the N\textsubscript{2}O\textsubscript{5} production caused, as before, by the delayed production of NO\textsubscript{2}. On the other hand, at 0\,slm the density profile of N\textsubscript{2}O\textsubscript{5} does not follow the density profile of NO\textsubscript{2}. Rather, the N\textsubscript{2}O\textsubscript{5} density decreases as the discharge transitions into the NO\textsubscript{x} mode, as its formation requires both NO\textsubscript{2} and NO\textsubscript{3}. In the NO\textsubscript{x} mode, the density of NO\textsubscript{3} remains low (not shown), and acts to limit the formation of N\textsubscript{2}O\textsubscript{5} even when the NO\textsubscript{2} density is large. \\

Considering that the model was developed with a focus on the production and consumption dynamics of O\textsubscript{3}, through the generation of NO by R9 as the main pathway, it can be stated that the model accurately simulates O\textsubscript{3} during the SDBD operation, fulfilling its initial objective. On the other hand, it is observed that, for low flow rates, the simulation of other species shows greater differences, but illustrates the general behavior of those densities. At high flow rates, the simulation deviates further from the experimental measurements. This indicates weaknesses in the model when the role of vibrationally excited nitrogen is decreased at higher flow rates. It should also be noted that the determined input parameters were mainly tailored to achieve a good match between measured and simulated O\textsubscript{3} densities. Attempts at tuning these parameters towards achieving better agreement for NO\textsubscript{2} and N\textsubscript{2}O\textsubscript{5}, however, generally lead a worse overall agreement. In addition to the chemistry, an improvement of the way in which gas transport is treated in the model may also be lead to improved results at higher flow rates. Overall, the current model is capable of producing accurate dynamics for O\textsubscript{3}, but requires further improvements to achieve good agreement with the dynamics of NO\textsubscript{x} production across the mode transition, especially at higher flow rates. \\

\addtolength{\textheight}{2.7pt}
\renewcommand{\arraystretch}{1.2}
\begin{table*}[htbp]
    \caption{System of heavy particle collisions considered in the zero-dimensional model for the determination of the temporal evolutions of the listed reactive species. Energy units: Gas temperature \textit{T}\textsubscript{g} [K]. Rate coeff. units: Single body reaction [s\textsuperscript{-1}]; Two-body reaction [m\textsuperscript{3}s\textsuperscript{-1}]; Three-body reaction [m\textsuperscript{6}s\textsuperscript{-1}]. M represents a third body and is given by the neutral gas number density as determined by the ideal gas law. Where no \textit{T}\textsubscript{g} dependence is given for a reaction, values refer to those determined at 300\,K. \\}
    \begin{tabular}{@{}p{0.052\textwidth}p{0.42\textwidth}p{0.39\textwidth}c@{}}
        \toprule
        No. & Reaction & Rate coefficient & Ref. \\
        \midrule
        \multicolumn{4}{l}{\emph{Atomic oxygen reactions}} \\
        R1 & \ce{O + O + M -> O2 + M} & $4.5 \times 10^{-46}e^{630/T_\mathrm{g}}$ & \cite{herron2001} \\
        R2 & \ce{O + O2 + M -> O3 + M} & $6.0 \times 10^{-46}(T_\mathrm{g}/300)^{-2.6}$ & \cite{iupac_atkinson} \\
        R3 & \ce{O + O3 -> O2 + O2} & $8.0 \times 10^{-18}e^{-2060/T_\mathrm{g}}$ & \cite{iupac_atkinson} \\
        R4 & \ce{O + NO + M -> NO2 + M} & $f(T_\mathrm{g})$ & \cite{nist_herron}$^{a}$\\
        R5 & \ce{O + NO2 -> NO + O2} & $5.1 \times 10^{-18}e^{198/T_\mathrm{g}}$ & \cite{iupac_atkinson} \\
        R6 & \ce{O + NO2 + M -> NO3 + M} & $f(T_\mathrm{g})$ & \cite{iupac_atkinson}$^{a}$\\
        R7 & \ce{O + NO3 -> O2 + NO2} & $1.7 \times 10^{-17}$ & \cite{iupac_atkinson} \\
        R8 & \ce{O + N2(A^3\Sigma) -> NO + N(^2D)} & $7.0 \times 10^{-18}$ & \cite{capitelli2000} \\
        R9 & \ce{O + N2(\nu) -> NO + N} & $1.0 \times 10^{-17}$ & \cite{capitelli2000} \\
        \multicolumn{4}{l}{\emph{Atomic nitrogen reactions}} \\
        R10 & \ce{N + N + M -> N2 + M} & $8.3 \times 10^{-46}(500/T_\mathrm{g})$ & \cite{herron2001} \\
        R11 & \ce{N + NO -> N2 + O} & $2.1 \times 10^{-17}(100/T_\mathrm{g})$ & \cite{herron2001,JPL19_5} \\
        R12 & \ce{N + NO2 -> N2O + O} & $5.8 \times 10^{-18}(220/T_\mathrm{g})$ & \cite{herron2001,JPL19_5} \\
        R13 & \ce{N + NO3 -> NO + NO2} & $3.0 \times 10^{-18}$ & \cite{herron2001} \\
        R14 & \ce{N + O + M -> NO + M} & $6.3 \times 10^{-45}(140/T_\mathrm{g})$ & \cite{herron2001}$^{b}$ \\
        R15 & \ce{N + O2 -> NO + O} & $3.3 \times 10^{-18}e^{-3150/T_\mathrm{g}}$ & \cite{JPL19_5} \\
        R16 & \ce{N + O3 -> NO + O2} & $1.0 \times 10^{-22}$ & \cite{JPL19_5} \\
        R17 & \ce{0 -> N} & $10^{18}$  & $^{c}$ \\
        \multicolumn{4}{l}{\emph{Ozone and nitrogen oxide reactions}} \\
        R18 & \ce{O3 + NO -> NO2 + O2} & $2.07 \times 10^{-18}e^{-1400/T_\mathrm{g}}$ & \cite{iupac_atkinson} \\
        R19 & \ce{O3 + NO2 -> NO3 + O2} & $1.4 \times 10^{-19}e^{-2470/T_\mathrm{g}}$ & \cite{iupac_atkinson} \\
        R20 & \ce{O3 + M -> O + O2 + M} & $9.61 \times 10^{-16}e^{-11600/T_\mathrm{g}}$ & \cite{jones1962} \\
        R21 & \ce{O3 + O2(a^1\Delta) -> O + O2 + O2} & $5.2 \times 10^{-17}e^{-2840/T_\mathrm{g}}$ & \cite{herron2001,iupac_atkinson} \\

        R22 & \ce{O3 + N2(A^3\Sigma) -> N2 + O2 + O} & $3.36 \times 10^{-17}$ & \cite{nist_herron} \\
        R23 & \ce{O3 + N2(A^3\Sigma) -> NO + NO + O} & $8.4 \times 10^{-18}$ & \cite{nist_herron} \\
        R24 & \ce{O3 + NO3 -> NO2 + O2 + O2} & $1.0 \times 10^{-23}$ & \cite{hjorth1992}$^{b}$ \\
        
        R25 & \ce{NO + NO2 + M -> N2O3 + M} & $f(T_\mathrm{g})$ & \cite{iupac_atkinson}$^{a}$\\
        R26 & \ce{NO + NO3 -> NO2 + NO2} & $1.8 \times 10^{-17}e^{110/T_\mathrm{g}}$ & \cite{iupac_atkinson} \\
        R27 & \ce{NO2 + NO2 + M -> N2O4 + M} & $f(T_\mathrm{g})$  & \cite{iupac_atkinson}$^{a}$\\
        R28 & \ce{NO2 + NO3 + M -> N2O5 + M} & $f(T_\mathrm{g})$ & \cite{iupac_atkinson}$^{a}$\\
        R29 & \ce{NO2 + NO3 -> NO2 + NO + O2} & $4.35 \times 10^{-20}e^{-1335/T_\mathrm{g}}$ & \cite{iupac_atkinson} \\
        R30 & \ce{NO3 + NO3 -> NO2 + NO2 + O2} & $8.5 \times 10^{-19}e^{-2450/T_\mathrm{g}}$ & \cite{JPL19_5} \\
        R31 & \ce{NO3 -> NO + O2} & $2.6 \times 10^{6}e^{-6100/T_\mathrm{g}}$ & \cite{johnston1986} \\
        R32 & \ce{N2O3 + M -> NO + NO2 + M} & $f(T_\mathrm{g})$ & \cite{iupac_atkinson}$^{a}$ \\
        R33 & \ce{N2O4 + M -> NO2 + NO2 + M} & $f(T_\mathrm{g})$ & \cite{iupac_atkinson}$^{a}$ \\
        R34 & \ce{N2O5 + M -> NO2 + NO3 + M} & $f(T_\mathrm{g})$ & \cite{iupac_atkinson}$^{a}$ \\
        R35 & \ce{N2O + N2(A^3\Sigma) -> O + N2 + N2} & $9.3 \times 10^{-18}e^{-120/T_\mathrm{g}}$ & \cite{nist_herron} \\
        R36 & \ce{N2O + N2(A^3\Sigma) -> N2 + N2 + O} & $3.3 \times 10^{-18}e^{-120/T_\mathrm{g}}$ & \cite{herron2001} \\
        R37 & \ce{NO + NO + O2 -> NO2 + NO2} & $4.25 \times 10^{-51}e^{663.5/T_\mathrm{g}}$  & \cite{iupac_atkinson} \\        
        \bottomrule
    \end{tabular} \\[0.375em]
    \footnotesize\raggedright
    $^{a}$ Calculated using the analytical expression given in the cited reference. \\
    $^{b}$ Rate coefficient estimated in the cited reference. \\
    $^{c}$ Production rate in m\textsuperscript{-3}s\textsuperscript{-1}, estimated by comparison of measured and simulated density profiles. This reaction is a general source \\ of N, included here as a proxy for processes such as electron impact dissociation of N\textsubscript{2}, for example.
    \label{tab:reactions}
\end{table*}

\section{Conclusion and future work}
\label{chap:conclusion}
The densities of O\textsubscript{3}, NO\textsubscript{2} and N\textsubscript{2}O\textsubscript{5} were measured in an SDBD reactor, operated with dry synthetic air as the feed gas of the system, with flow rate variations from 0\,slm to 10\,slm. At low flow rates, two distinct operating modes were observed: O\textsubscript{3} was the dominant species in the first mode, while NO\textsubscript{2} dominated the second. This transition between modes has been demonstrated in other works before, which our work builds upon \cite{shimizu2012, park2018}. At higher flow rates, the mode transition was not clearly observed, highlighting the necessity of the vibrational state of nitrogen, N\textsubscript{2}(\textit{v}) with \textit{v}\,$\geq$\,12 to drive the mode transition. The increase in the flow rate impacts the residence time and appears to reduce the vibrational temperature, leading to lower densities of N\textsubscript{2}($v>12$). \\

A chemical kinetics model was developed to enhance the understanding of the processes behind the discharge dynamics. The model simulates the densities of the species involved in the reactions, within the dry synthetic air mixture. The results point to the importance of N\textsubscript{2}($v>12$), and its influence on the mode transition. As part of the code development, parameters such as gas temperature \textit{T}\textsubscript{g}, steady-state vibrational temperature \textit{T}\textsubscript{v}, and vibrational time constant $\tau$\textsubscript{v} were defined. It was observed that these parameters directly affect the steady-state values and the initial rise times of the considered reactive species. \\

The measured results were compared with the simulated values. The agreement obtained for O\textsubscript{3} was sufficiently accurate to conclude that the presented reactions and mechanisms describe the main processes involved in O\textsubscript{3} production and consumption, during the discharge operation. In the case of NO\textsubscript{2}, the agreement was acceptable but insufficient to fully describe the process, with each increase in flow rate leading to greater discrepancies with the measured data. Specifically, for the densities of NO\textsubscript{2} and N\textsubscript{2}O\textsubscript{5} a potentially significant role for N atoms was proposed, which requires further study. \\

Finally, it is possible to conclude that the model achieves an adequate representation of the experimental behavior of this type of SDBD discharge. However, the focus in this work was on the production and consumption dynamics of O\textsubscript{3}, and its dependency on the highly excited vibrational state of N\textsubscript{2}, N\textsubscript{2}($v>12$). Likewise, the importance of dissociative production of N on the production of nitrogen oxides was proposed, presenting a potential approach for future works to achieve a more complete description of the entire system.

\begin{acknowledgments}
    This study was funded by the German Research Foundation (DFG) with the Collaborative Research Centre CRC1316 ``Transient atmospheric plasmas: from plasmas to liquids to solids" (projects A7 and A9) and by the JSPS Core-to-Core Program ``Data Driven Plasma Science".
\end{acknowledgments}

\section*{Data Availability Statement}

\begin{center}
    \renewcommand\arraystretch{1.2}
    \begin{tabular}{| >{\raggedright\arraybackslash}p{0.39\linewidth} | >{\raggedright\arraybackslash}p{0.54\linewidth} |}
        \hline
        \textbf{AVAILABILITY OF DATA} & \textbf{STATEMENT OF DATA AVAILABILITY}\\  
        \hline
        Data openly available in a public repository that does not issue DOIs
        &
        The data that support the findings of this study will be openly available at \href{https://rdpcidat.rub.de/}{https://rdpcidat.rub.de/}.\;\;\;\;\;
        A reference number will be assigned after finalization of the peer review process.
        \\\hline
    \end{tabular}
\end{center}

\clearpage
\section*{References}
\vspace{-1em}
\bibliography{aipsamp}

\end{document}